\documentclass[12pt,preprint]{aastex}
\shorttitle{Gravitational Microlensing by Rotating Stars}
\shortauthors{Habib Ebrahimnejad Rahbari, Mohammad Nouri-Zonoz and
Sohrab Rahvar }
\begin{document}

\title{Gravitational Microlensing by Rotating Stars}


\author{Habib Ebrahimnejad Rahbari\altaffilmark{1}, Mohammad Nouri-Zonoz  \altaffilmark{1,3}  and Sohrab Rahvar \altaffilmark{2,3}}
\email{nouri@khayam.ut.ac.ir}
\email{rahvar@sharif.edu}


\altaffiltext{1}{Department of Physics, University of Tehran,
North Karegar St., Tehran 14395-547, Iran}
\altaffiltext{2}{Department
of Physics, Sharif University of Technology,
 P.O.Box 11365--9161, Tehran, Iran}
\altaffiltext{3}{Institute for Studies in Theoretical Physics and
Mathematics, P.O.Box 19395--5531, Tehran, Iran}

\begin{abstract}
Using astrometry of microlensing events we study the effect of
angular momentum as compared to that of the parallax. For a
rotating lens it is shown that the effect of the angular momentum
deviates the center of images from that in the simple standard
microlensing. This effect could be observed by future astrometry
missions such as GAIA and SIM for lenses with angular momentum
$S\gtrsim \times 10^{48} kg m^2 sec^{-1}$. It is shown that for
extreme black hole lenses the corresponding mass should be more
than $10^3 M_{\odot}$.
%Lenses with the corresponding angular momentum
%We show how the angular momentum detection is sensitive to the
%lens parameters and argue that low mass lenses with a given
%angular momentum have a higher chance to be detected.
\end{abstract}

\keywords{gravitation -- stars: rotation -- cosmology:
gravitational lensing}

\section{Introduction}
Gravitational lensing and microlensing have been treated and
employed as powerful tools to study astrophysical objects and
phenomena (Schnierder, Ehlers and Falco 1993; Mollerach, Esteban
2002). Among many examples and applications one could mention the
search for MACHO candidates at the galactic halo using
gravitational microlensing with the recent data pointing at the
absence of them \cite{mil05}. Other application of gravitational
microlensing include (a) study of stellar atmosphere (Cassan et
al. 2004; Abe et al. 2003; Bryce et al. 2002), (b) planet
searching (Bennet et al. 1999) ,(c) exotic matter searches such as
magnetic masses \cite{rah03,nou97,lyn98} and wormholes
(Safonova, Torres and Romero 2002), (d) possible black hole
detection based on gravitational microlensing method (Agol et. al
2002). Furthermore it seems that future observations will be based
on astrometric microlensing which is hoped to improve previous
studies. Another application which is the concern of this paper
includes angular momentum detection by photometric and astrometric
microlensing. This in turn could enhance detection of black hole
candidates through this extra parameter. \\
The outline of the paper is as follows: In Section
\ref{photometry} we introduce photometric and astrometric
microlensing. Section \ref{kerr} contains gravitational
microlensing in kerr metric and sensitivity of angular momentum detection to the
lens parameters. Conclusion and summery are given in Section \ref{sum}.
%%%%%%%%%%%%%%%%%%%%%%%%%%%%%%%%%%%%%%%%%%%%%%%%%
\section{Photometric and Astrometric Microlensing}\label{photometry}
Simple microlensing events occur when the approximation of a
point-like deflector and point-like source in a relative uniform
motion is valid in the Galactic scale \cite{pac86}. At a given
time t, the light magnification $A(t)$ of a point-like source
located at lens-source $D_{ls}$ and observer-lens $D_{ol}$
distances induced by a point-like deflector of mass $M$ is given
by:
\begin{equation}
A(t)={u^2(t)+2\over u(t)\sqrt{4+u^2(t)}},
\end{equation}
where $u(t)$ is the impact parameter (distance between lens and
source in the lens plane), expressed in units of the "Einstein
Radius" $R_{E}$ and is given by:
\begin{equation}
R_{E} = \sqrt{{4GDM \over c^2 }}\;\;\;\; , \;\;\;\; D =
{D_{ol}D_{ls}\over D_{os}},
\end{equation}
where $G$ is the Newtonian gravitational constant and $c$ is the
velocity of light. Assuming a source moving at a constant relative
transverse speed $v_{T}$, reaching its minimum distance $u_{0}$ to
the lens in the lens plane at time $t_{0}$, $u(t)$ is given by:
\begin{equation}
u(t) = \sqrt{u_0^2 + ({t-t_0 \over t_E})^2}\;\;\;\; , \;\;\;\; t_E
= {R_E\over v_T},
\end{equation}
where $t_E$, the "Einstein crossing time", is the only measurable
parameter providing useful information about the lens in the
approximation of the simple microlensing. Within this
approximation, the light-curve of a microlensed star is fully
determined by the parameters $F_b,u_0,t_0$ and $t_E$. In the next
two subsections we introduce astrometry and parallax effect
and their signature in the gravitational microlensing.
\subsection{Astrometric microlensing in Schwarzschild space time}
For a point like source at distance $u={b\over R_E}$ in the lens
plane from a point like gravitational lens, it is well known that
the images are located at:
\begin{equation}
{u_{\pm}^{I}={u\pm\sqrt{u^2+4}\over2}}.
\end{equation}
Corresponding amplifications are given by:
\begin{equation}
A_{\pm}={1\over2}(1\pm{2+u^2\over u\sqrt{u^2+4}}).
\end{equation}
The location of the center of images with respect to lens, in the lens plane, is
defined by:
\begin{equation}
{\bf R}_{image}={{{\bf u}_{+}^{I}|A_{+}|+{\bf u}_{-}^{I}|A_{-}|}\over
{|A_{+}|+|A_{-}|}}.
\end{equation}
We are interested in the deflection of the combined images, i.e.
the location of centroid with respect to the source, so we
transform ${\bf R}_{image}$ to ${\bf R^{'}}_{image}$ i.e :
\begin{equation}
{\bf R^{'}}_{image}={\bf R}_{image}-{\bf u}\;\;\; then \;\;\;
{\bf R^{'}}_{image}={u\over u^2+2}\frac{{\bf u}}{u}. \label{c_image}
\end{equation}
As $u$ changes with time ${\bf R^{'}}_{image}$ plots a pattern. In
the case of the Schwarzschild space-time the pattern is an ellipse. \\
To study the asymptotic behavior of the standard astrometric
microlensing, we examine equation (\ref{c_image}). If
$u{\rightarrow}\infty$, the centroid shift falls off like $1\over
u$ , this can be compared with the photometric amplification A,
which falls off like: $1\over u^4$. These asymptotic results
illustrate one of the important differences between astrometric
and photometric microlensing namely that the {\it centroid shift
falls off much more slowly than amplification}. In consequence,
cross section for astrometric events is much larger than that for
photometric events \cite{mir96,dom00,hon01}.
%%%%%%%%%%%%%%%%%%%%%%%%%%%%%%%%%%%%%%%%%%%%%%%%%%%%%
\subsection{Parallax effect in microlensing}
\begin{figure}
%\epsscale{0.9}
%\includegraphics[angle=0,scale=0.72]{f1.eps}
%\plotone{light_curve.eps}
\plotone{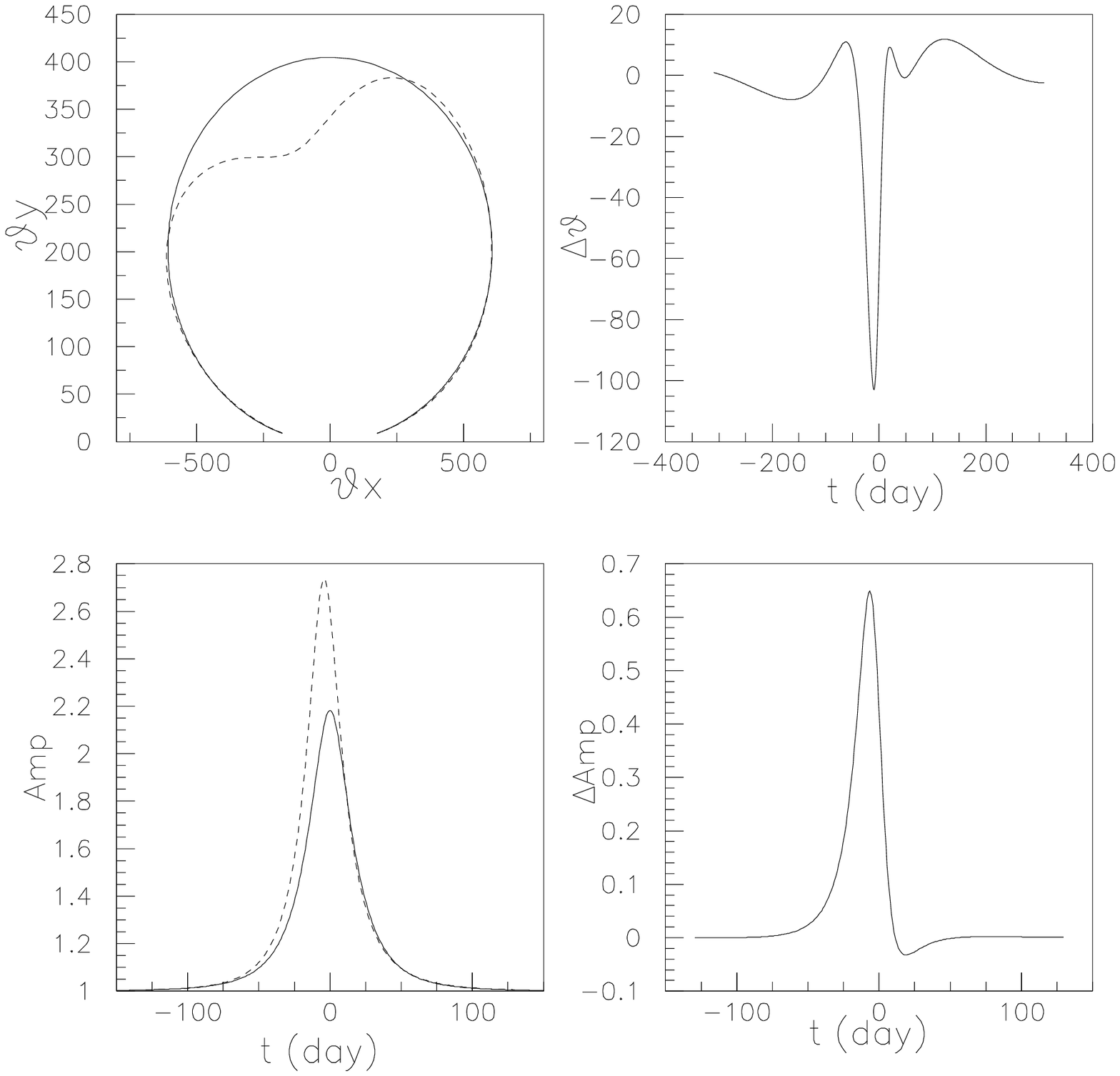} \caption{ Astrometry and Photometry of a
microlensing event in schwarzschild metric. The simple
microlensing event (solid-line) is compared with considering the
parallax effect (dashed-line). Top-left panel shows the astrometry
of center of images for both cases and at top-right panel the
contrast between them is shown. The lower-left panel is the
photometry and the lower-right panel is the contrast between them.
The parameters of lens is chosen as $t_e = 40$ days, $t_0 = 0$,
$u_0 = 0.5 $, $M= M_{\odot}$, $D_{os}=10 kpc$ and $D_{ls}= 5 kpc$.
\label{f1}}
\end{figure}

If the variation of the Earth's velocity around the sun is not
negligible with respect to the projected transverse speed of the
deflector, then the apparent trajectory of the deflector with
respect to the line of sight is a cycloid instead of a straight
line. The resulting amplification versus time is therefore
affected by the so-called {\it parallax effect}. This effect is
more easily observable for long duration events (several months),
for which the change in the Earth`s velocity is important
\cite{alc95}. If ${\bf u}_D(t)$ is the position of the deflector
in the deflector's transverse plane and ${\bf u}_E(t)$ is the
intercept of the Earth-source line of sight with this plane, then:
\begin{equation}
{u(t)=\left|{\bf u}_{D}(t)-{\bf u}_{E}(t)\right|^{1\over2}},
\end{equation}
where
\begin{equation}
 {\bf u}_D(t)=({t-t_0\over t_E}cos\theta-u_0sin\theta){\bf
\hat{i}}+ ({t-t_0\over t_E}sin\theta-u_0cos\theta){\bf \hat{j}},
\end{equation}
and $\theta$ is the angle between the projected lens trajectory
and the projected major axis of the Earth's orbit in the deflector
plane. Here $u_0$ is the closest approach of the lens to the sun
at the lens plane. Neglecting the Earth's orbital eccentricity,
${\bf u}_E(t)$ is given by:
\begin{equation}
{{\bf u}_E(t)={\delta u}sin(\xi(t)){\bf \hat{i}}-{\delta
u}{cos(\xi(t))}{cos(\beta}){\bf \hat{j}}},
\end{equation}
where $\delta u={{a_\oplus (1-x)}/R_{E}} $ is the projection of
the Earth's orbital radius in the deflector plane in unit of
Einstein radius and $x=\frac{D_{ol}}{D_{os}}$, $\beta$ is the
angle between the ecliptic and deflector planes and $\xi(t)$ is
the phase of the Earth relative to its position at ${\bf u}_E
={\delta u}cos(\beta){\bf \hat{j}}$. The distortion of the light
curve is important if the Earth's orbital velocity around the sun
is not negligible with respect to the projected transverse speed
of the deflector:
\begin{equation}
{\tilde{v}}={R_E\over t_E(1-x)}={a_{\oplus}\over \delta ut_E}
\end{equation}
Rahvar et al. (2003) proposed an observation strategy, using the
alert and follow-up telescopes for that can put constrain on the
mass and distance of lens in the direction of Magellanic Clouds.
Adding a third telescope such as GAIA or SIM one can break the
degeneracy between the parameters of lens through astrometric
measurements \cite{rah05}. Fig. \ref{f1} shows the photometry and
the astrometry of center of images for the case of simple and
considering the parallax effects for a typical lens located at $5
kpc$ from the observer with the source star at $10 kpc$.
%%%%%%%%%%%%%%%%%%%%%%%%%%%%%%%%%%%%%%%%%%%%
\section{Microlensing in Kerr space}
\label{kerr} Gravitational lensing in the field of a rotating star
has been studied previously, for example in
\cite{bra86,iba83,gli99}. Here we employ the formalism
introduced by Ibanez (1983) in which the first approximation in $G$
of Kerr metric was used. In the slow-motion or the fast-motion approximation,
the bending angle is given by:
\begin{equation}
\label{dev_eq}
 \Delta {\bf n} =
-\frac{2}{c}\int^{+\infty}_{-\infty}\nabla \phi dt +\frac{4G}{c^3
b^2}[\frac{2}{b^2} {\bf b}.({\bf n_i} \times {\bf S}){\bf b} -{\bf
(n_i\times S)}],
\end{equation}
where ${\bf b}$ is the impact parameter, ${\bf S}$ spin of
the lens and ${\bf n_i}$ a unit vector representing the initial
direction of incoming photons toward the lens. We note that for
${\bf S}$ normal to the lens plane, the second term in the right hand side of
equation (\ref{dev_eq}) vanishes and we can not observe the signature of
angular momentum on the position of images. The relation between
the position of image and source at the lens plane perpendicular
to the line of sight is given by:
%which is more appropriate for our study
% But the very useful formalism for our
%work is the Ibanez's article.
%He studied this problem using the Linearized General Theory of
%Relativity. The equation of the lens [J.Ibanez] is given by:
\begin{eqnarray}
\label{proj1} x_0 &=& x + {{R_E}^2\over x_0^2+y_0^2}x_0 +
{2R^2\over (x_0^2+y_0^2)^2}x^2_0
- {R^2\over x_0^2+y_0^2}. \\
\label{proj2}
 y_0 &=& y + {{R_E}^2\over x_0^2+y_0^2}y_0 + {2R^2\over
(x_0^2+y_0^2)^2}x_0y_0,
\end{eqnarray}
where $R^2 = {4GS D\over c^3}$ and $S$ is the
projected angular momentum on the lens plane. $(x,y)$ and $(x_0, y_0)$ are the
components of the source and image positions respectively. (Y-axis is chosen
along projection of spin at the lens plane). At the lens plane,
the ratio of the terms corresponding to that of the angular momentum to the
of mass, in the deviation of light ray is $R^2/{{R_E}^2 x_0}=S/c/{M
x_0}=(v/c)({R_g}/{x_0})$, where $v$ is the rotation speed of lens,
$R_g$ is gyration radius and $x_0$ is the position of images. Sine
$v<<c$ and $R_g<<x_0$, the ratio of the spin to the mass term
should be smaller than one, means that $S<<M x_0 c$. For a typical
lens of solar mass, in the Milky Way, the images are produced at
an astronomical distance from the lens, in other words $S<< 10^{50}
kg m^2/sec$. For lenses with larger masses and located at the
galactic scale, the upper limit on the spin will be higher. \\
For a given source, equations (\ref{proj1}) and (\ref{proj2}) in
principle have three solutions, which means that rotating stars produce
three images through lensing where the third image is very close to
the lens. Glicenstein (1999) has showed that the third image
having a very small impact parameter, it always eclipsed by the lens
itself. Using equations (\ref{proj1}) and (\ref{proj2}), the
amplification of the light of an image produced by a rotating lens is
given by:

\begin{equation}
A =  (1- {{R_E}^4\over b^4} - {4R^4\over b^6} -
{4R^2{R_E}^2x_0\over b^6})^{-1}
\end{equation}

% Extending the above considerations to a more general case,
%we calculate equation of lens and amplification of light when {\bf
%S} acquires a general direction i.e ${\bf S } = S\cos\chi {\bf i}
%+ S\sin\chi {\bf j}$ where $\chi$ is the angle between the spin
%angular momentum vector and x-axis. Then the equation of lens is
%obtained:
%$$x_0 = x + {Q\over x_0^2+y_0^2}x_0 + {{2Rsin(\chi-\varphi)}\over (x_0^2+y_0^2)^{3\over 2}}x^2_0 - {{Rsin\chi}\over x_0^2+y_0^2}\eqno(16-1)$$
%$$y_0 = y + {Q\over
%x_0^2+y_0^2}y_0 + {{2Rsin(\chi-\varphi)}\over (x_0^2+y_0^2)^{3\over 2}}x_0y_0 +{{Rcos\chi}\over x_0^2+y_0^2}\eqno(16-2)$$
%And the amplification of light is the jacopian of transformation of $(x,y)$ to $(x_0,y_0)$  then is obtained:
%$${A = {1\over 1- {Q^2\over b^4} - {4R^2\over b^6} - {4RQsin(\chi-\varphi)\over b^5}}}\eqno(17)$$
%This means that if the spin of lens is normal to the lens plane
%then we could not detect it by a microlensing experiment and there
%is degeneracy in the direction of spin normal to the lens plane.
%-Fig. 3 -Fig.4
%%%%%%%%%%%%%%%%%%%%%%%%%%%%%%%%%%%%%%%%%%%%%%%%%%%%%%%
\subsection {Astrometric microlensing in Kerr space}
\begin{figure}
%\epsscale{0.9}
%\includegraphics[angle=0,scale=0.72]{f2.eps}
%\plotone{light_curve.eps}
\plotone{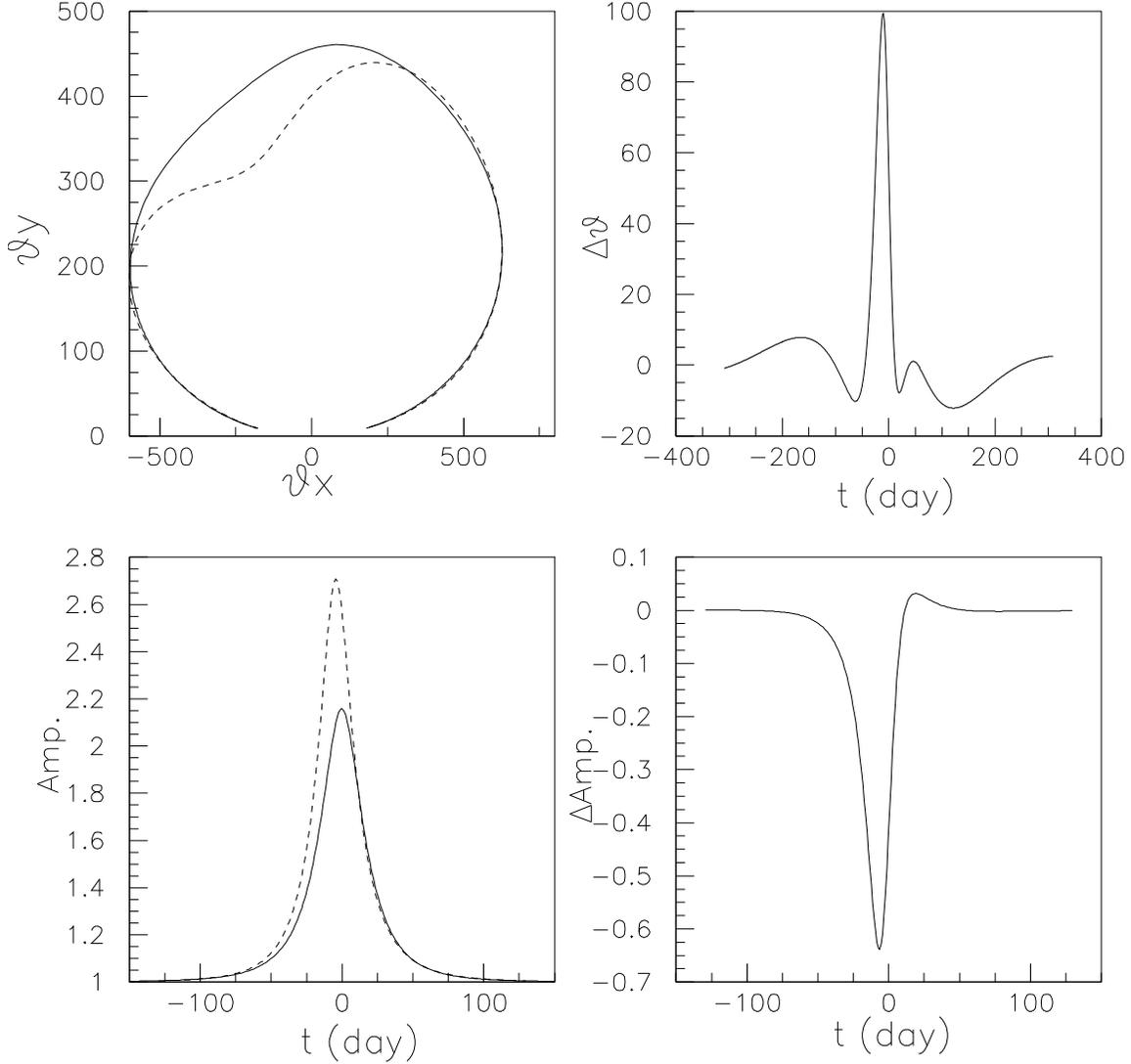} \caption{ Astrometry and Photometry of a
microlensing event in Kerr metric. The microlensing event for the
static observer (solid-line) is compared with that of considering
the parallax effect (dashed-line). Top-left panel shows the
astrometry of center of images for both cases and top-right panel
shows the contrast between them. The lower-left panel is the
photometry and the lower-right panel is the contrast between them.
The parameters of lens is chosen as $t_e = 40$ days, $t_0 = 0$
days, $u_0 = 0.5 $, $M= M_{\odot}$, $D_{os}=10 kpc$ and $D_{ls}= 5
kpc$ and spine is taken as ${\bf S}=(-cos{\pi\over 3}{\bf
i}+sin{\pi\over 3}{\bf j})\times 10^{41}Kgm^2s^{-1}$}
 \label{f2}
 \end{figure}

\begin{figure}
\plotone{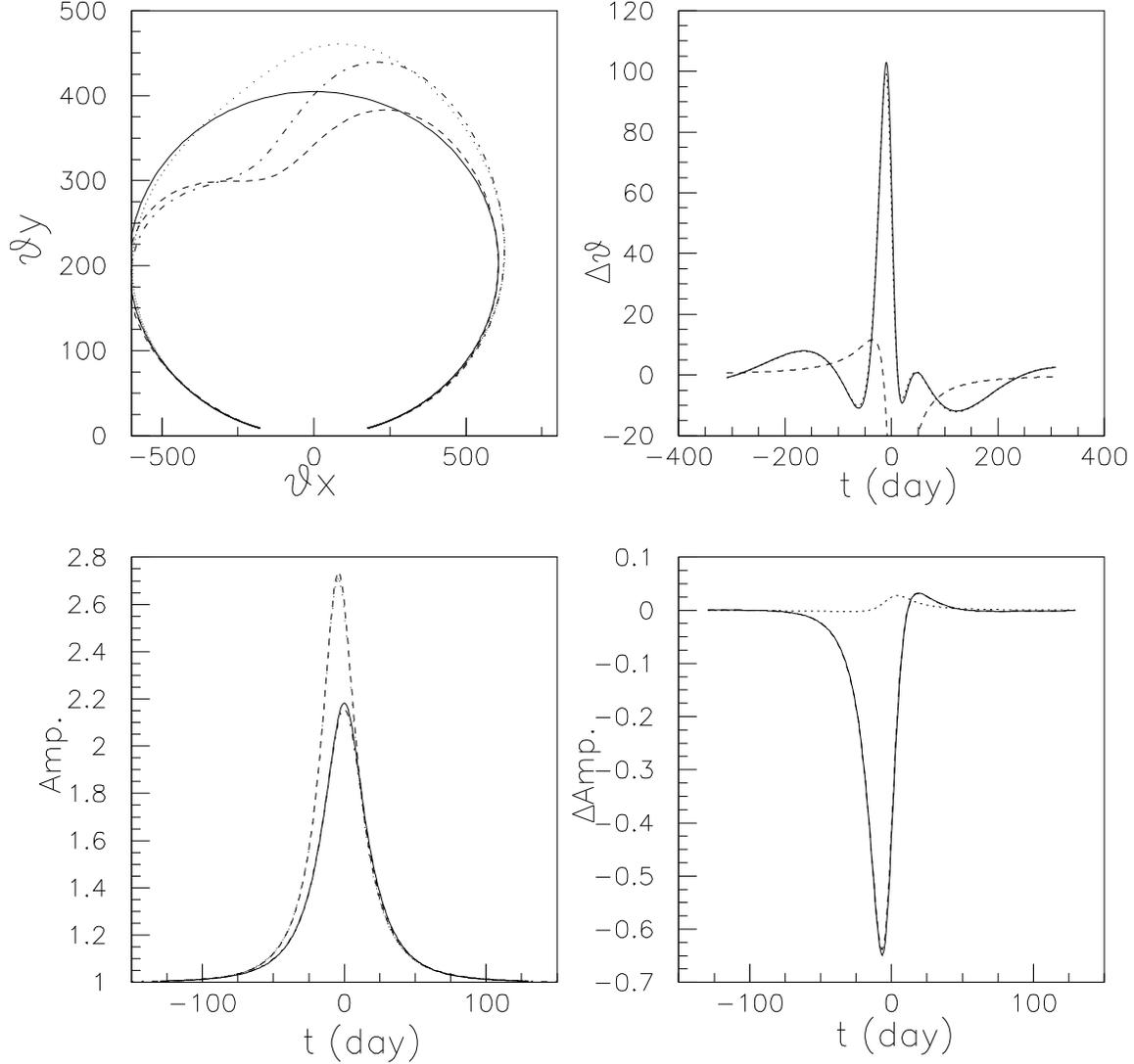}
 \caption{ Comparison of the astrometry and photometry in the
Schwarzschild and Kerr metrics. The parameters of lens are as in
Figs. (\ref{f1}) and (\ref{f2}). At the top-left hand side panel,
astrometry for the static observer in the Schwarzschild metric
(solid-line), considering the parallax effect (dashed-line), in
Kerr metric with static observer (dotted-line) and with the
parallax effect (dashed-dotted line) is presented. The top-right
panel shows the difference between the position of center of
images for Schwarzschild and Schwarzschild-parallax (solid-line),
Kerr and Kerr-Parallax (dotted-line) which in coincide to the fist
curve and difference between the Kerr and Schwarzschild metrics
(dashed-line). The lower plane is the photometry with an without
parallax effects for Schwarzschild (Solid-line), with parallax
(dashed-dotted line), Kerr (dashed-line) and Kerr with parallax
(dotted line). At the right-down panel, the difference between the
Schwarzschild and Schwarzschild-parallax (solid-line) and Kerr and
Kerr-Parallax (dashed-line) which in coincide to the fist curve
and difference between the Kerr and Schwarzschild metrics (dotted
line). } \label{trig_proj} \label{f3}
\end{figure}

\begin{figure}
\plotone{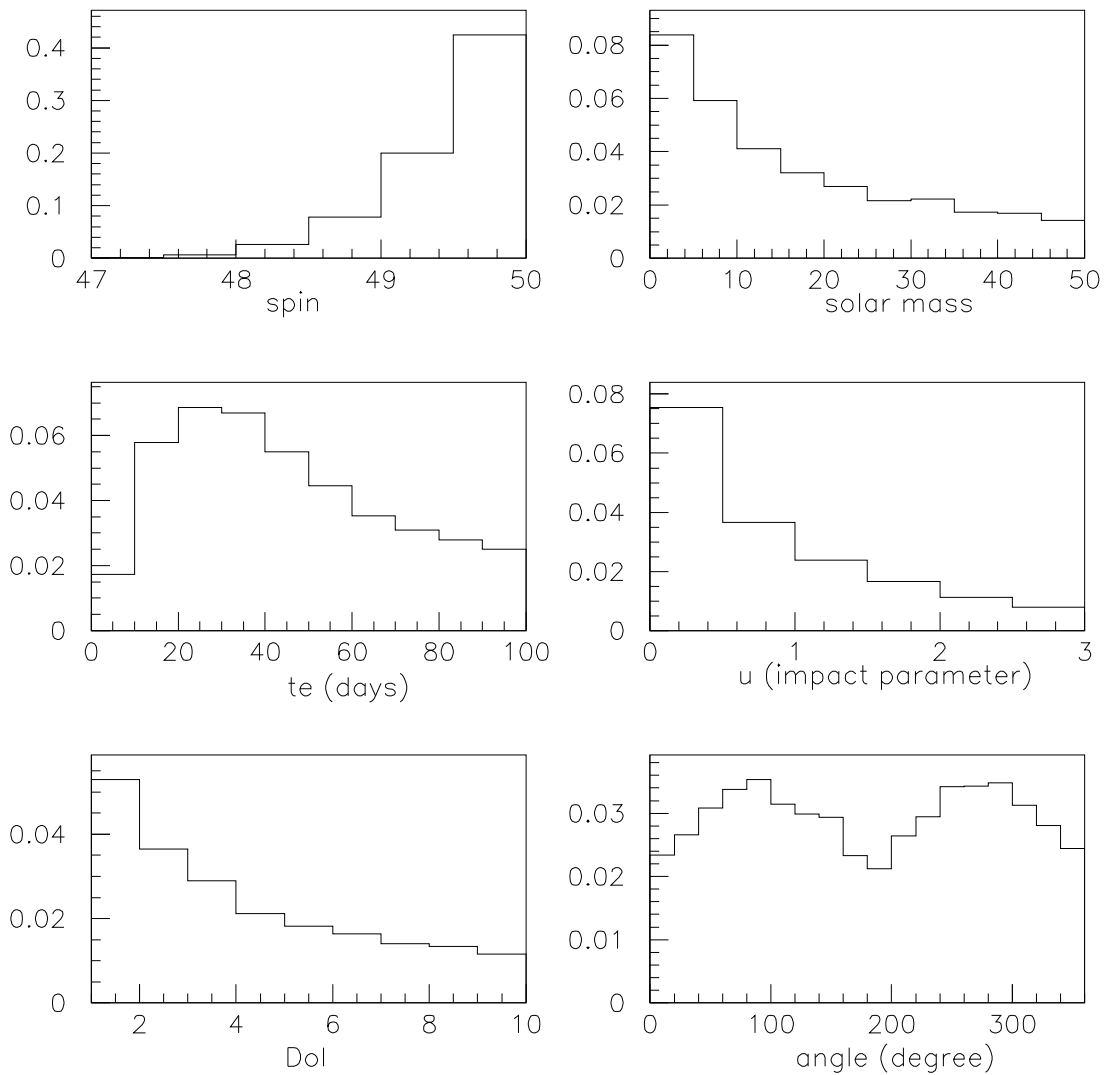} \caption{Fraction of lenses that can show the
signature of spin for the telescope with $10 \mu as$ angular
resolution. The parameters of the lens, the spin, mass, Einstein
crossing time, impact parameter, observer-lens distance and the
orientation of spin with respect to the observer-lens-source plane
are generated uniformly to show the sensibility of them to the
rotating star detection.} \label{f4}
\end{figure}

As the equations (\ref{proj1}) and (\ref{proj2}) are too complex
to be solved analytically, we employ the method of initial
guess and substitute the schwarzschild solution in (\ref{proj1})
and (\ref{proj2}) and through iteration obtain the location of images
in the linear Kerr metric. Fig. \ref{f2} shows the astrometry of the center of images and the photometry of
a microlensing event by a rotating star. Similar to the case of
Schwarzschild metric the parallax effect alters the path of center
of images in the Kerr metric. Astrometries in
Schwarzschild and Kerr metrics with and without the parallax effect are
compared in Fig. \ref{f3}. Since the parallax is not an intrinsic effect
it alters the center of images with the same amount both in Kerr and
Schwarzschild spaces, Fig .\ref{f3}. Subtraction of the parallax effect from the
path of the center of the images may help us to distinguish the Kerr from
Schwarzschild metric.\\
%%%%%%%%%%%%%%%%%%%%%%%%%%%%%%%%%%%%%%%%%%%%%%
%\section{Possibility of angular momentum detection though astrometry}\label{detection}
In practice, the astrometric measurements can be
done by a ground-based telescope accompanying a space-based
interferometry telescope. The ground-based telescope signals the
microlensing events and undertakes photometric measurement and the
space-based telescope measures the displacement of image
centroids. We examine the sensitivity of angular momentum
detection as a function of lens parameters. To do so we need
a Monte Carlo simulation generating the mass function as well as the corresponding
angular momentum for a given direction of Milky Way and use
observational efficiency to estimate the number of
events. However, since the dependence of angular momentum to the
mass function of stars is not well know, we generate a uniform
distribution of lens parameters and use the criterion that the difference between
the astrometry of Kerr and ordinary lenses should be more than the
angular resolution of GAIA or SIM ($10 \mu as>$). The fraction
of events with the detectable angular momentum is shown in Fig.
\ref{f4}. It is seen that the angular momentum detection is correlated to
 all of the parameters of the lens. Increasing the spin, in contrast to the
 mass rises the detection probability. Very short and long Einstein crossing times
 are not favored for the spin detection but on the other hand small impact parameter is more convenient for our propose.
  The distance of lens from the observer also decreases the
sensibility of angular momentum detection. The effect of distance decreases the displacement
 of center of images, Fig. \ref{f4}.

\section{Conclusion}\label{sum}
In the present study we examined the possibility of angular momentum detection of
rotating lenses by the gravitational microlensing. Out of
two methods of photometric and astrometric microlensing, the latter is
more feasible. Future space-based interferometry
telescopes such as GAIA and SIM can be used for this purpose. Studying the
detection sensitivity in terms of the lens parameters showed that the centriod of images
in the lenses with angular momentum larger than $S\gtrsim \times 10^{48} kg m^2 sec^{-1}$ deviates more
 than $10 \mu a$ from that of the Schwarzschild case. Recent analysis of
X-ray spectral data from {\it ASCA} and {\it RXTE} of the two black hole
 candidates GRO J1655-40 and 4U 1543-47 has estimated the corresponding dimensionless
  spin parameter $a_{*} = (Sc/GM^2)$, $\sim  0.7 - 0.75$ and $\sim 0.85-0.9$ respectively \cite{sha05}.
   Considering our  estimation of the spin detection limit implies that the mass of an extreme black
    hole $(a_{*} = 1)$ with observable angular momentum should be more than $10^3 M_{\odot}$.

\section*{Acknowledgments}
The authors would like to thank  S.E. Vazquez, O. Wucknitz, L.
Wisotzkiand  and C. Fendt for their usefull comments and for
providing some of the references.  M-NZ  and  H.E also thank
University of Tehran for supporting this project under the grants
provided by the research council.

%\section{Appendicial material}

%\clearpage

%\begin{figure}
%\plotone{trig_eff.eps} \caption{ This figure shows two dimensional
%trigger efficiency in terms of events duration and $R$. It is seen
%that for long duration events the chance of alert if hight than
%the short term events. \label{trig_eff}}
%\end{figure}

\end{document}